\documentclass[12pt]{article}
\usepackage{epsf,latexsym}
\usepackage{amsmath,amsfonts,amssymb,amsthm} 
\usepackage[applemac]{inputenc}
\epsfverbosetrue
\newtheorem{thm}{Theorem}[section]

\theoremstyle{definition}                    
\newtheorem{defn}[thm]{Definition}
\theoremstyle{remark}

\numberwithin{equation}{section}  

\usepackage[ps,arc,frame]{xy}

\newcommand{\rxy}[1]{{\begin{xy}
0;<2mm,0mm>:<0mm,2mm>::0;0,#1\end{xy}}}

\newcommand{\rxya}[2]{{\begin{xy} 0;<2mm,0mm>:<0mm,2mm>::0;0,
,(5,-2)*{a}
,(10,-2)*{b}
,(2,-5)*{a}
,(2,-10)*{b}
,(5,-5)*\cir(#1,0){}
,(10,-5)*\cir(#1,0){}
,(5,-10)*\cir(#1,0){}
,(10,-10)*\cir(#1,0){}
#2\end{xy}}}

\newcommand{\rxyg}[2]{{\begin{xy} 0;<2mm,0mm>:<0mm,2mm>::0;0,
,(5,-2)*{a} ,(10,-1.8)*{b} ,(15,-2)*{c} ,(20,-1.8)*{d} ,(2,-5)*{a}
,(1.8,-10)*{b} ,(2,-15)*{c} ,(1.8,-20)*{d} ,(5,-5)*\cir(#1,0){}
,(10,-5)*\cir(#1,0){} ,(15,-5)*\cir(#1,0){} ,(20,-5)*\cir(#1,0){}
,(5,-10)*\cir(#1,0){} ,(10,-10)*\cir(#1,0){} ,(15,-10)*\cir(#1,0){}
,(20,-10)*\cir(#1,0){} ,(5,-15)*\cir(#1,0){} ,(10,-15)*\cir(#1,0){}
,(15,-15)*\cir(#1,0){} ,(20,-15)*\cir(#1,0){} ,(5,-20)*\cir(#1,0){}
,(10,-20)*\cir(#1,0){} ,(15,-20)*\cir(#1,0){} ,(20,-20)*\cir(#1,0){}
#2\end{xy}}}

\epsfverbosetrue
\newcommand{\double}[1]{\mathbb{#1}}

\newcommand{\cc}{\double{C}}

\newcommand{\rr}{\double{R}}
\newcommand{\zz}{\double{Z}}

\newcommand{\kk}{\double{K}}

\newcommand{\Aa}{\mathcal{A}}

\newcommand{\hhh}{\double{H}}

\newcommand{\dd}{\mathcal{D}}

\newcommand{\hh}{\mathcal{H}}

\newcommand{\T}{{\rm tr}}

\newcommand{\op}{\oplus}

\newcommand{\bb}{\begin{eqnarray}}
\newcommand{\ee}{\end{eqnarray}}
\newcommand{\eee}{\nonumber\end{eqnarray}}
\newcommand{\pp}[1]{\begin{pmatrix} #1 \end{pmatrix}}
\newcommand{\qq}{\quad}

\begin{document}

\font\twelve=cmbx10 at 13pt
\font\eightrm=cmr8

\thispagestyle{empty}

\begin{center}

CENTRE DE PHYSIQUE TH\'EORIQUE \footnote{\,  Unit\'e Mixte de Recherche
(UMR 6207)
 du CNRS  et des Universit\'es Aix--Marseille 1 et 2\\
${}$\qq\qq\qq et  Sud
Toulon--Var, Laboratoire affili\'e \`a la FRUMAM (FR 2291)}
\\ CNRS--Luminy, Case 907\\ 13288 Marseille Cedex 9\\
FRANCE\\

\vspace{2cm}

{\Large\textbf{On a Classification of Irreducible \\
 Almost-Commutative
Geometries IV}} \\

\vspace{1.5cm}

{\large Jan--Hendrik Jureit
\footnote{\, and Universit\'e de Provence and Universit\"at Kiel,\\
${}$\qq\qq\qq\ jureit@cpt.univ-mrs.fr},
Christoph A. Stephan
\footnote{\, and Universit\'e de Provence and Universit\"at Kiel,\\
${}$\qq\qq\qq\ stephan@cpt.univ-mrs.fr} }

\vspace{1.5cm}

{\large\textbf{Abstract}}
\end{center}
In this paper we will classify the finite spectral triples with $KO$-dimension six,
following the classification found in \cite{1,2,3,4}, with up to four
summands in the matrix algebra. Again, heavy use is
made of Krajewski diagrams \cite{Kraj}. 
This work has been inspired
by the recent paper by Alain Connes \cite{connes6} and John Barrett
\cite{barrett6}.

In the classification we find that the standard model of particle physics in its 
minimal version
fits the axioms of noncommutative geometry in the case of $KO$-dimension
six. By minimal version it is  meant that at least
one neutrino has to be massless and mass-terms mixing particles
and antiparticles are prohibited.
\vspace{1.5cm}

\vskip 1truecm

\noindent CPT-P75-2006 \\
PACS-92: 11.15 Gauge field theories\\
\indent MSC-91: 81T13 Yang-Mills and other gauge theories

\vskip 1truecm


\vspace{2cm}

\section{Introduction}

Recently Alain Connes \cite{connes6} and John Barrett \cite{barrett6} proposed
to change the $KO$-dimension for the finite part of almost-commutative
spectral triples from zero to six. Based on this assumption they constructed
a version of the standard model of particle physics which allowed for
right-handed massive neutrinos in every generation of fermions and 
a Majorana-mass resulting in the see-saw-mechanism. 
Furthermore the long standing problem of fermion-doubling could be cured.
In the case of $KO$-dimension six it is possible to directly project out the superfluous 
degrees of freedom, as is shown in detail in \cite{barrett6}.

The price which
had to be paid is that not all the axioms of noncommutative geometry \cite{book,grav}
are satisfied by this model, notably the orientability axiom which 
fails on the Lepton-sector \cite{connespriv}. Also the Poincar\'e duality
needs to be modified in the sense that the leptonic sector and the quark sector
provide two separate generators of K-homology. Each of these sectors fulfils the
Poincar\'e duality \cite{connespriv}.

In this paper we will assume that all the axioms of noncommutative
geometry hold and classify the corresponding finite spectral triples
following \cite{1,2,3,4}. This classification is based on the classification
of finite spectral triples of Mario Paschke, Andrzej Sitarz and Thomas
Krajewski \cite{Pasch, Kraj}. The main tool used to find the possible
spectral triples are Krajewski diagrams \cite{Kraj} which have already
been used in \cite{1,2,3,4}. Passing from $KO$-dimension zero
to $KO$-dimension six implies a few changes in the definition of
a real, finite spectral triple. It will be shown in detail that every real,
finite spectral triple in $KO$-dimension six 
still allows to split the Hilbert space into left- and right-handed particle
and antiparticle subspaces, although this choice may be ambiguous 
and it does not in principle forbid lepto-quark-like mass terms. 

Imposing the axioms of noncommutative geometry will lead us
to seven possible minimal Krajewski diagrams, two of which contain
the first family of the standard model of particle physics in its minimal version. Thus,
if one requires all the axioms to hold, one has to abandon Majorana-masses
for right-handed neutrinos and at least one neutrino has to remain
massless. This should be compared with the case in $KO$-dimension
zero \cite{1,2,3,4}, where 66 Krajewski diagrams appeared, all corresponding to noncommutative geometries which obey to the axioms. It is quite
remarkable that a reduction in the input, i.e. no $S^0$-real structure has
to be assumed,  does lead to a reduction in the number of possible geometries.

A version of the standard model with four summands
in the matrix algebra, with right-handed
neutrinos and Majorana-masses has been treated in detail in \cite{ko6}.
This necessitated a modification of the axioms of noncommutative
geometry, notably the orientability axiom.

\section{Basic Definitions}

In this classification we are interested in real, finite spectral triples
with $KO$-dimension six and metric dimension zero, \cite{real,grav}.
The metric dimension being zero follows from the requirement of
finiteness since this implies that the internal Dirac operator has
only a finite number of eigenvalues. 
Note that no $S^0$-real structure is imposed. We will show that
the axioms still allow to split the Hilbert space into left-handed and 
right-handed particle and anti-particle subspaces and that the $S^0$-real
structure, which is needed in $KO$-dimension zero to perform this split, 
follows  naturally from the axioms. It follows also that the 
Dirac operator must not contain mass terms which connect particles
to antiparticles. 

\begin{defn} A real, finite spectral triple of $KO$-dimension six is given by
($\Aa,\hh,\dd, $ $J,\chi$)  with a finite dimensional real algebra
$\Aa $, a faithful representation
$\rho$ of $\Aa$ on  a finite dimensional complex Hilbert space $\hh$. Three
additional operators are defined on
$\hh$: the Dirac operator
$\dd$ is selfadjoint, the real structure $J$ is antiunitary,  and the chirality
$\chi$ which is an unitary involution. These operators  satisfy:
\bb
\hspace{-0.1cm}\bullet \hspace{0.5cm} J^2= \chi^2 =1,\qq
[J,\dd]=\{ J,\chi\}=0,\qq\dd\chi =-\chi \dd, \qq
[\chi,\rho(a)]=0, \cr [\rho(a),J\rho(b)J^{-1}]=
[[\dd,\rho(a)],J\rho(b)J^{-1}]=0,  \forall a,b \in \Aa.
\label{commu}
\ee

Note that in $KO$-dimension six the commutator $[J,\chi]=0$ from 
$KO$-dimension zero becomes an anti-commutator \cite{connes6, barrett6}.
\\ \\
$\bullet$ The chirality
 can be written as a finite sum $\chi =\sum_i\rho(a_i)J\rho(b_i)J^{-1}.$
This condition is called orientability. The finite sum is a zero 
dimensional Hochschild cycle.
\\ \\
$\bullet$ The intersection form
$\cap_{ij}:=\T(\chi \,\rho (p_i) J \rho (p_j) J^{-1})$ is non-degenerate,
$\rm{det}\,\cap\not=0$. The
$p_i$ are minimal rank projections in $\Aa$. This condition is called
{\it Poincar\'e duality}.
\end{defn}

The algebra is a finite sum of $N$ simple
algebras , $\Aa = \oplus_{i=1}^N \Aa_i$
with  $\Aa_i=M_{n_i}(\kk_i)$,
 and $\kk_i=\rr,\cc,\hhh$ where $\hhh$
denotes the quaternions.
\\ \\
We will now give a derivation of the substructure of the operators
and the Hilbert space constituting the spectral triple. 
It will turn out that the vanishing anti-commutator $\{ J,\chi\}=0$ and
the axiom of orientability replace the $S^0$-real structure of the case with
$KO$-dimension zero. 

\paragraph{The chirality, the real structure and the Hilbert space:}
Since the chirality is a unitary operator with $\chi^2 = 1$ 
we can build  the projectors $(1\mp \chi )/2$  which allow to  decompose
the Hilbert space
\bb
\hh=\hh^-\op\hh^+ \, . \label{espacedehilbert}
\ee
The first
component  corresponds in physics to left-handed particles and to  charge conjugate 
right-handed particles, $\chi = -1$, the second component
corresponds to right-handed particles and the charge conjugate of left-handed particles, $\chi = +1$. 
The real structure
anti-commutes with the chirality,  $\{ J,\chi \}=0$, therefore it maps the  
subspace $\hh^-$ to the  subspace $\hh^+$ and vice versa. 
And since
$J^2=1$ we have dim$\hh^-$=dim$\hh^+=n$.  As a convention we will 
take the basis of $\hh$ in which the chirality is a diagonal matrix with
eigenvalues $\mp 1$
\bb
\chi = (- 1_n) \oplus 1_n
\ee
and in which the real structure takes the standard form
\bb
J = \pp{ 0 & 1_n \\ 1_n & 0 } \circ {\rm complex} \; {\rm conjugation}.
\ee

\paragraph{The representation:}
The chirality and the real structure automatically require the representation $\rho$
of the algebra $\Aa$
to be block-diagonal on the subspaces $\hh^-$ and $\hh^+$
\bb
\rho(a) = \rho^-(a) \oplus \rho^+(a), \; \; a\in \Aa.
\label{repsplit}
\ee
Concerning the algebra $\Aa = \oplus_{i=1}^N \Aa_i$, we restrict ourselves to the  easy case, $\kk=\rr,
\hhh$ in all components $\Aa_i=M_{n_i}(\kk_i)$ of the algebra. The sub-algebras 
$M_n(\rr)$ and
$M_n(\hhh)$ only have  one irreducible representation, the fundamental
one on $\cc^{(n)}$, where $(n)=n$ for $\kk=\rr$ and
$(n)=2n$ for $\kk=\hhh$. All of the following arguments also hold for $\kk = \cc$,
but the notation becomes more opaque since the complex conjugate of
the fundamental representation has to be taken into account.
We notice that the axiom $ [\rho(a),J\rho(b)J^{-1}]=0$
for all $a,b \in \mathcal{A}$ requires $\rho ( a )$
to be of the form
\bb
\rho ( a ) = \left( \oplus_{i,j=1}^N \rho^-(a_i,a_j) \right) \oplus  
\left(\oplus_{i,j=1}^N \rho^+(a_i,a_j) \right)
\label{kranot}
\ee 
where $\rho^-$ and $\rho^+$ can be either of the form
\bb
\rho^-(a_i,a_j) :=  a_i \otimes 1_{m_{ij}}
\otimes 1_{n_j} \quad
\rho^+(a_i,a_j) := 1_{n_i} \otimes 1_{m_{ij}} \otimes
a_j 
\label{standard}
\ee
or of the form
\bb
\rho^-(a_i,a_j) :=  1_{n_i} \otimes 1_{m_{ij}} \otimes
a_j \quad
\rho^+(a_i,a_j) :=  a_i \otimes 1_{m_{ij}}
\otimes 1_{n_j}. 
\ee
The multiplicities $m_{ij}$ are non-negative integers and we denote by
$1_n$ the $n\times n$ identity matrix and set by convention $1_0:=0$.
Algebra elements $a_i$ are taken to be from he $i$th summand 
$M_{n_i}(\kk_i)$ of the algebra $\Aa = \oplus_{i=1}^{N} M_{n_i}(\kk_i)$.

The real structure $J$ permutes the two main
summands $\rho^-$ and $\rho^+$ of $\rho$ (\ref{kranot})
and complex conjugates them.  The chirality reads explicitly
\bb
\chi&=&(\oplus_{i,j=1}^N 1_{n_i} \otimes (-1) 1_{m_{ij}} \otimes
1_{n_j})\
\op\ (\oplus_{i,j=1}^N 1_{n_i} \otimes 1_{m_{ij}} \otimes
1_{n_j}).
\label{chi}
\ee
To simplify the following calculations we will also assume $m_{ij}=1$
in all sub-representations. 

The orientability axiom allows to  put further restrictions on the sub-representations
$\rho(a_i,a_j) = \rho^-(a_i,a_j) \oplus \rho^+(a_i,a_j)$. Let $\chi_{ij}$ 
and $J_{ij}$ be the
chirality and the real structure restricted to the subspace $\hh_{ij} = \hh_{ij}^- \oplus
\hh_{ij}^+$ corresponding to $\rho(a_i,a_j)$. Then the orientability axiom
requires that there exist $a_{i,r},b_{i,r} \in \Aa_i$ and  $a_{j,r},b_{j,r} \in \Aa_j$
such that
\bb
\sum_r \rho(a_{i,r},a_{j,r}) J_{ij} \rho(b_{i,r},b_{j,r}) J_{ij}^{-1} \, = \, \chi_{ij}.
\ee
Without loss of generality we can choose the sub-representation to be of
the form \ref{standard} and find in a straightforward calculation
\bb
\sum_r \rho(a_{i,r},a_{j,r}) J_{ij} \rho(b_{i,r},b_{j,r}) J_{ij}^{-1} \, = 
\sum_r (a_{i,r} \otimes b_{j,r} ) \oplus (b_{i,r} \otimes a_{j,r})
\ee
which is required to be equal to $\chi_{ij} = (-1)1_{(n_i +n_j)} \oplus 1_{(n_i+n_j)}$.
It follows immediately that $i \neq j$. The same argument holds for the complex 
case, $\kk = \cc$. 

Note that this argument forbids Majorana masses for right-handed neutrinos
if the orientability axiom is used in its classical form \cite{ko6}.

\paragraph{The Dirac operator:}
The Dirac operator anti-commutes with the chirality, $\{ \dd, \chi \} =0$,
and maps therefore the sub-space $\hh^-$ to $\hh^+$ and vice versa.
Taking into account the self-adjointness of $\dd$ leads to the following
general form
\bb
\dd = \pp{ 0 & \Delta \\ \Delta^* & 0}, \;\; {\rm with} \; \; \Delta:\hh^- \rightarrow \hh^+.
\ee
Furthermore $\dd$ commutes with the real structure $J$,
\bb
[\dd,J] = \pp{ \overline{\Delta}^*-\Delta & 0 \\ 0 & \overline{\Delta} - \Delta^*} = 0
\ee
and therefore $\Delta^t = \Delta$. And so the Dirac operator reads:
\bb
\dd = \pp{ 0 & \Delta \\ \overline{\Delta} & 0} \;\; {\rm with} \;\;  \Delta^t = \Delta.
\label{Diracop}
\ee
This form of the Dirac operator seems to deviate from the standard one
used in the literature, but this is only due to the choice of basis which  is
obtained by ordering the  particle and anti-particle multiplets with respect
to their chirality. 

\paragraph{The first order axiom:}
We will now employ the first order axiom, 
\bb
[[\dd,\rho(a)],J\rho(b)J^{-1}]=0,  \forall a,b \in \Aa, 
\ee
to put further restrictions on the fine structure of the Dirac operator
and the representation. Using the equation \ref{Diracop} for the 
Dirac operator and the representation \ref{repsplit} one finds with
\bb
J \rho(a) J^{-1} =  \overline{\rho}^{\,-}(a) \oplus \overline{\rho}^{\,+}(a)
\ee
after a simple calculation the following two equations for the first order axiom
\bb
\Delta \, \rho^-(a) \,\overline{\rho}^{\,+}(b) -  \rho^+(a)\, \Delta \, \rho^+(b)
-  \overline{\rho}^{\,-}(b) \, \Delta \, \rho^-(a) + \overline{\rho}^{\,-}(b)  \,
\rho^+(a) \, \Delta = 0  \label{firstorder} \\
\overline{\Delta} \, \rho^+(a) \,\overline{\rho}^{\,-}(b) -  
\rho^-(a)\, \overline{\Delta} \, \rho^-(b)
-  \overline{\rho}^{\,+}(b) \, \overline{\Delta} \, \rho^+(a) + \overline{\rho}^{\,+}(b)  \,
\rho^-(a) \, \overline{\Delta} = 0,
\ee
for all $a,b \in \Aa$.
One notices immediately that these two equations are equivalent and we will therefore
use from now on the first equation \ref{firstorder}.

To determine the fine structure of $\Delta$ we first restrict ourselves to
the subspace $\hh_{ij} = \hh_{ij}^- \oplus \hh_{ij}^+$ so that
\bb
M_{ij} := \Delta |_{ \hh_{ij}^-} \, :  \hh_{ij}^- \longrightarrow  \hh_{ij}^+\, .
\ee
Without loss of generality we choose the representation restricted to
$\hh_{ij}$ to be 
\bb
\rho(a_i,a_j) = a_i \otimes 1_{n_j} \oplus  1_{n_i} \otimes a_j,
\ee
which corresponds to the choice of equation (\ref{standard}) with
multiplicity $1_{m_{ij}}= 1$.
For the restricted Dirac operator, real structure and chirality we choose 
the following:
\bb
\dd_{ij} = \pp{ 0 & M_{ij} \\ \overline{M_{ij}} & 0}, \;\; 
J_{ij} = \pp{0 & 1_{n_i+n_j}  \\ 1_{n_i+n_j} & 0} \circ {\rm cc}, \; \;
\chi_{ij} =\pp{(-1)1_{n_i+n_j} & 0 \\ 0& 1_{n_i+n_j}}.
\ee
Keeping in mind that we are for reasons of simplicity dealing with real
and quaternionic matrix algebras, the first order axiom in the form of equation 
\ref{firstorder}  reads 
\bb
M_{ij} \, (b_i \otimes a_j) - (a_i \otimes 1_{n_j}) \, M_{ij} \, (b_i \otimes 1_{n_j})
- (1_{n_i} \otimes b_j) \, M_{ij} \, (1_{n_i} \otimes a_j) + (a_i \otimes b_j) \, M_{ij} = 0
\nonumber \\
\label{firstorder2}
\ee
for all $a_i,b_i \in \Aa_i$ and $a_j,b_j \in \Aa_j$ with $i\neq j$. In particular we can
therefore  choose $a_i = b_i = 0$ and $a_j = b_j = 1_{n_j}$
which leads to $M_{ij} = 0$. 

As a first result from this analysis we conclude therefore that the Dirac operator
cannot map subspace to each other which are also mapped to each other
by the real structure. This is a direct consequence of the orientability axiom
which requires $i\neq j$ in \ref{firstorder2}. The same argument holds
for the complex case and for any multiplicity $m_{ij}$ in the representation
\ref{kranot}.

Let us now consider the more general case where we restrict ourselves
to the subspace 
\bb
\tilde{\hh} = \hh_{ij}^- \oplus \hh_{rs}^- \oplus \hh_{ij}^+ \oplus \hh_{rs}^+
\label{hh2}
\ee
with the obvious restrictions of the real structure and the chirality. The
sub-matrix $\Delta$ of the Dirac operator $\dd$ restricted to $\tilde{\hh}$
is  
\bb
\Delta |_{\tilde{\hh}} = \pp{0 & M_{ij}^{rs} \\ M_{rs}^{ij} & 0},
\ee
with
\bb
M_{ij}^{rs} : \hh_{ij}^- \rightarrow \hh_{rs}^+ \quad {\rm and} \quad
M_{rs}^{ij} : \hh_{rs}^- \rightarrow  \hh_{ij}^+.
\ee
From $\Delta = \Delta^t$ follows $(M_{ij}^{rs})^t = M_{rs}^{ij}$ and thus
\bb
\Delta |_{\tilde{\hh}} = \pp{0 & M_{ij}^{rs} \\ (M_{ij}^{rs})^t & 0} = \pp{0 & M \\ M^t & 0},
\label{delta2}
\ee
where we have dropped the indices to simplify the notation.

We have now to distinguish two classes of possible representations
on $\tilde{\hh}$. The first possible representation restricted to $\tilde{\hh}$
is given by
\bb
\rho(a) |_{\tilde{\hh}} = \rho^-(a_i,a_j) \oplus \rho^-(a_r,a_s) \oplus
 \rho^+(a_i,a_j) \oplus \rho^+(a_r,a_s)
\ee
with
\bb
 \rho^-(a_i,a_j) = a_i \otimes 1_{n_j}, \quad \rho^+(a_i,a_j) = 1_{n_i} \otimes a_j,
 \nonumber \\
\rho^-(a_r,a_s) = a_r \otimes 1_{n_s}, \quad \rho^+(a_r,a_s) = 1_{n_r} \otimes a_s, 
\ee
with $i\neq j$ and $r\neq s$.  Plugging the representation and $\Delta$ into
the first order condition \ref{firstorder} we find for the sub-matrix $M$
the following equation
\bb
M \, (b_r \otimes a_s) - (a_i \otimes 1_{n_j}) \, M \, (b_r \otimes 1_{n_s})
- (1_{n_i} \otimes b_j) \, M \, (1_{n_r} \otimes a_s) + (a_i \otimes b_j) \, M = 0.
\label{firstorder3}
\ee
This equation has two solutions with $M\neq 0$:  $i=s$ with $\Aa_i = \Aa_j= \rr,\cc$ and
 $r=j$ with $\Aa_r=\Aa_j = \rr,\cc$. The possibilities $i,j,r,s$ pair-wise different and
 $i=r$, $j=s$ lead to $M=0$. 
 
We have therefore the two possible representations with $M\neq 0$:
\bb
 \rho^-(a_i,a_j) = a_i 1_{n_j}, \quad \rho^+(a_i,a_j) =  a_j,
 \nonumber \\
\rho^-(a_r,a_i) = a_r, \quad \rho^+(a_r,a_i) = a_i 1_{n_r}, 
\ee 
for $i=s$ and
\bb
 \rho^-(a_i,a_j) = a_i, \quad \rho^+(a_i,a_j) =  a_j 1_{n_i},
 \nonumber \\
\rho^-(a_j,a_s) = a_j 1_{n_s}, \quad \rho^+(a_j,a_s) =  a_s, 
\ee
for $r=j$. These two possibilities are obviously equivalent.
\\ \\
The second general class of representations is of the form 
\bb
 \rho^-(a_i,a_j) = a_i \otimes 1_{n_j}, \quad \rho^+(a_i,a_j) = 1_{n_i} \otimes a_j,
 \nonumber \\
\quad \rho^-(a_r,a_s) = 1_{n_r} \otimes a_s,  \quad \rho^+(a_r,a_s) = a_r \otimes 1_{n_s}.
\ee
Using again the first order axiom \ref{firstorder} we find 
\bb
M \, (a_r \otimes b_s) - (a_i \otimes 1_{n_j}) \, M \, (1_{n_r} \otimes b_s)
- (1_{n_i} \otimes b_j) \, M \, (a_r \otimes 1_{n_s}) + (a_i \otimes b_j) \, M = 0.
\label{firstorder4}
\ee
For $i=r$ ($\Aa_i=\Aa_j$) this has a general solution of the form $M= 1_{n_i} \otimes m$
with $m \in M_{n_j\times n_s}(\cc)$ and for $j=s$ ($\Aa_j=\Aa_s$) the solution has the form 
$M= m \otimes 1_{n_j}$  with $m \in M_{n_i\times n_r}(\cc)$. One has to
keep in mind that $i\neq j$ and $r\neq s$ are required from the orientability
axiom. All other possible equalities for $i,j,r,s$ lead again to $M=0$.

We have therefore again two possible representations with $M\neq 0$:
\bb
 \rho^-(a_i,a_j) = a_i \otimes 1_{n_j}, \quad \rho^+(a_i,a_j) = 1_{n_i} \otimes a_j,
 \nonumber \\
\quad \rho^-(a_i,a_s) = 1_{n_i} \otimes a_s,  \quad \rho^+(a_i,a_s) = a_i \otimes 1_{n_s}
\ee
for $i=r$ and
\bb
 \rho^-(a_i,a_j) = a_i \otimes 1_{n_j}, \quad \rho^+(a_i,a_j) = 1_{n_i} \otimes a_j,
 \nonumber \\
\quad \rho^-(a_r,a_j) = 1_{n_r} \otimes a_j,  \quad \rho^+(a_r,a_j) = a_r \otimes 1_{n_j}.
\ee
for $j=s$. These considerations can be easily adapted to the complex case which
leads to some extra conditions which are treated in detail in \cite{Kraj}. Further
restrictions also emerge from the axiom of Poincar\'e duality. This will
be treated in the section on Krajewski diagrams.

The restriction to the subspace $\tilde{\hh}$ it is now straightforward to 
choose particle- and antiparticle-subspaces $\hh^P$ and $\hh	^A$. We define
\bb
\hh^P := \hh_{ij}^- \oplus \hh_{rs}^+ \quad {\rm and} \quad
\hh^A := \hh_{ij}^+ \oplus \hh_{rs}^-,
\ee
and identify $ \hh_{ij}^{-(+)}$ with the left-handed (anti)particles and $ \hh_{rs}^{+(-)}$
with the right-handed (anti)particles. In this basis the Dirac operator takes
the well  known form 
\bb
\dd = \pp{\Delta^P & 0 \\ 0 & \overline{\Delta^P}} \quad {\rm with} \quad
\Delta^P = \pp{ 0 & M \\ M^* & 0}.
\ee
This split extends naturally to  the whole Hilbert space but it is off course
ambiguous to choose which subspace constitutes the particle space
and which the antiparticle space. The main point is that this split is always
possible which was not the case in $KO$-dimension zero without an
$S^0$-real structure. The crucial difference to the case of $KO$-dimension 
zero with $S^0$-real structure is that in principle lepto-quark like 
mass terms may appear in the Dirac operator. These were forbidden
by the $S^0$-real structure. The ambiguity in  the choice of particle
and antiparticles existed also in $KO$-dimension zero due to a freedom of
choice for the  $S^0$-real structure.

Finally we would like to give the general form of the chirality after splitting
the Hilbert space into particle- and antiparticle-subspaces, 
$\hh = \hh^P \oplus \hh^A$:
\bb
\chi&=&(\oplus_{i,j=1}^N 1_{(n_i)} \otimes \chi_{ji}1_{m_{ji}} \otimes
1_{(n_j)})\
\op\ (\oplus_{i,j=1}^N 1_{(n_i)} \otimes (-\chi_{ji})1_{m_{ji}} \otimes
1_{(n_j)}),
\label{chi}
\ee
where $\chi_{ij}=\mp 1$ according to our previous convention on left-(right-)handed
subspaces. The real structure is invariant under this change of basis.

Comparing to the classification of finite spectral triples \cite{Pasch,Kraj} we
find ourselves in the position to adapt directly the arguments which  lead
to the diagrammatic approach of Krajewski. For the choice of particle-
and antiparticle subspaces we will use the conventions of \cite{1}.

\section{Irreducibility and Krajewski diagrams}

We are again dealing with irreducible spectral triples so let us
recall the basic definitions.

\subsection{Irreducibility}

\begin{defn}
 i) A spectral triple $(\Aa,\hh,\dd)$ is {\it degenerate} if the kernel of
$\dd$ contains a non-trivial subspace of the complex Hilbert space $\hh$
invariant under the representation $\rho$ on $\hh$ of the real algebra
 $\Aa$.  \\
 ii) A non-degenerate spectral triple $(\Aa,\hh,\dd)$ is {\it reducible} if
there is a proper subspace
$\hh_0\subset\hh$ invariant under the algebra $\rho(\Aa)$  such that
$(\Aa,\hh_0,\dd|_{\hh_0})$ is a non-degenerate spectral triple. If the
triple is real and even, we require  the subspace
$\hh_0$ to be also invariant under the real structure $J$ and under the chirality
$\chi $ such that the triple $(\Aa,\hh_0,\dd|_{\hh_0})$ is again real and even.
\end{defn}

Krajewski and Paschke \& Sitarz have classified all finite,  
real spectral triples \cite{Pasch,Kraj}. Let us summarize the basics of this
classification  using Krajewski's diagrammatic
language.

\subsection{Conventions and multiplicity matrices}

We will again only treat  the easy case, $\kk=\rr,
\hhh$ in all components of the algebra to emphasise the differences
with respect to the case of $KO$-dimension zero. For further
details on the complex case and on multiple arrows we 
refer to \cite{1}.
 
We define the {\it multiplicity matrix} $\mu \in M_N(\zz)$,
$N$ being the  number of summands in $\Aa$, such
that $\mu _{ij}:=\chi _{ij} \, m_{ij}$, with $m _{ij}$ being
the multiplicities of the representation (\ref{kranot}) and
$ \chi _{ij}$ the signs of the chirality (\ref{chi}). There are $N$ minimal projectors in
$\Aa$, each of the form
$p_i=0 \op  \cdots \oplus 0\op \rm{diag}(1_{(1)},0,...,0) \op 0\oplus \cdots
\op 0$. With respect to the basis $p_i$, the matrix of the
intersection form is $\cap = \mu - \mu ^T$, the relative minus
sign has again its origin in the anti-commutation relation of the
real structure $J$ and the chirality $\chi$. 

If both entries $\mu _{ij}$ and $\mu _{ji}$ of the multiplicity
matrix are non-zero, then they must have the opposite sign.
This has again to be contrasted with the case in $KO$-dimension
zero, where the same sign is required.

$\bullet$ Poincar\'e duality: The last condition to be satisfied by the
multiplicity matrix reflects the Poincar\'e duality
and requires the multiplicity matrix to obey  $\det(\mu - \mu ^T)\not=0$.
Since the intersection form is an anti-symmetric matrix, this readily restricts us to finite 
spectral triples with an even number of summands in the matrix
algebra.

$\bullet$ The Dirac operator: The components of the (internal) Dirac
operator are represented by horizontal or vertical lines connecting two
nonvanishing entries of opposite signs in the multiplicity matrix $\mu $
and we will orient them from plus to minus. Each arrow represents a
nonvanishing, complex submatrix in the Dirac operator: For instance
$\mu_{ij}$ can be linked to $\mu_{ik}$ by
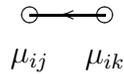
\begin{center}
\begin{tabular}{cc}
\rxy{
,(0,0)*\cir(0.7,0){}
,(5,0)*\cir(0.7,0){}
,(5,0);(0,0)**\dir{-}?(.6)*\dir{>}
,(0,-3)*{\mu_{ij}}
,(5,-3)*{\mu_{ik}}
}
\end{tabular} 
\end{center}
and this arrow represents respectively submatrices of $M$ in $\dd$ of
type $m\otimes 1_{(n_i)}$ with $m$ a complex $(n_j)\times(n_k)$ matrix.

\noindent Every arrow comes with three algebras:
Two algebras that localize its end
points, let us call them {\it right and left algebras}
and a third algebra that localizes the arrow, let us call it {\it colour
algebra}.  For the arrow presented above 
the left algebra is $\Aa _j$, the right algebra is $\Aa_k$ and the colour
algebra is $\Aa_i$.

We deduced  however that if $i=j$ or $k=j$ 
the corresponding spectral triple
does not satisfy the axiom of orientability, so the colour algebra
must not coincide with the left of the right algebra. Translated into the language
of Krajewski diagrams this means that the arrow must not touch the
diagonal of the diagram. 

\noindent The requirement of
non-degeneracy of a spectral triple means that every nonvanishing
entry in the multiplicity matrix
$\mu $ is touched by at least one arrow. We will also restrict ourselves
to minimal Krajewski diagrams. A minimal Krajewski diagram
is defined in detail in \cite{algo},
in short it means that it is not possible to remove an arrow from the diagram
without changing the multiplicity matrix.

$\bullet$ Convention for the diagrams: 
Our arrows always point from right chirality for particles and 
antiparticles, to 
left chirality for particles and antiparticles. 
As a further convention the horizontal arrows will encode particles 
and the vertical arrows encode antiparticles. This choice is of course
arbitrary. As in the case of the classification of finite spectral triples
of  $KO$-dimension zero \cite{1,2,3,4} there may appear "corners", i.e.
a horizontal arrow and a vertical arrow connected to a single point.
But since every arrow comes with its transposed arrow (through
the transposed multiplicity matrix), we can choose
here as well one pair of arrows to represent the particles and the other
to represent the antiparticles.

\section{The Classification}

As mentioned in the introduction, we will not give a complete derivation of
the physical content for the irreducible, minimal Krajewski diagrams under 
consideration.  For
this we refer to \cite{4}, where all the details of the resulting physical
models can be found. As we will see, the Krajewski diagrams of the
case with $KO$-dimension six form a subset of the diagrams found
in $KO$-dimension zero. Since the diagrams are taken to correspond
to irreducible spectral triples, only the first fermion family is contained
as a physical model. Further families have to be added by hand, these
spectral triples are no longer irreducible.

The minimal diagrams for the case of four summands in the matrix algebra
were computed with a computer program based on the algorithm 
presented in \cite{algo}. Only the calculation for the intersection form
was changed so that the condition det$(\mu - \mu^T)=0$ has to 
hold for the multiplicity matrix corresponding to the Krajewski diagram.
Thus also diagrams with arrows touching the diagonal appear in
the output of the algorithm and have to be removed by hand since
they do not satisfy 
the axiom of orientability.

In the classification in  \cite{1,2,3,4} all models were discarded 
that have either
\begin{itemize}\item
 a dynamically
degenerate fermionic mass spectrum,
\item
 Yang-Mills or gravitational
anomalies,
\item
 a fermion multiplet whose representation under
the little group is real or pseudo-real,
\item
or a massless fermion transforming
non-trivially under the little group.
\end{itemize}

Checking the Krajewski diagrams whether or not they meet
the above conditions is completely analogous to the $KO$-dimension
zero case. In fact the change of $KO$-dimension does not affect 
the representation, the Dirac operator or the physical models produced 
by a diagram. The only item of the spectral triple which is changed
is the chirality.

The multiplicity matrix is anti-symmetric so the Poincar\'e duality
can only be satisfied if the number of summands in the matrix
algebra is even.  The classification will be done for the cases
with two summands and four summands.  A classification beyond 
four summands is currently in progress.

\subsection{Two Summands}

In the case of two summands only one minimal Krajewski diagram exists:

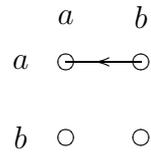
\begin{center}
\begin{tabular}{c}
\rxya{0.7}{
,(10,-5);(5,-5)**\dir{-}?(.6)*\dir{>}
}   \\ \\
\end{tabular}
\end{center}
Since the arrow touches the diagonal the  diagram cannot
represent a spectral triple which obeys to  the orientability axiom.
Therefore it will be discarded.

\subsection{Four Summands}

We will now analyse the diagrams produced by the computer program. 
They form a subset of the
Krajewski diagrams obtained for the four-summand case 
in $KO$-dim. zero treated in \cite{4}. This result is far from
obvious, for one might expect that new diagrams are possible
due to the sign change in the multiplicity matrix. As mentioned above 
the physical models
obtained from a given diagram do not depend on the $KO$-dimension.
We will therefore just give the  number of
the corresponding diagram in \cite{4} to simplify the comparison
for the reader. Furthermore we will quickly summarise the main results
from \cite{4} for each diagram and give the reason why they fail to
meet the classification requirements or, if they survive which physical
models they produce. 
\vspace{1\baselineskip}

{\bf Diagram 1:} 

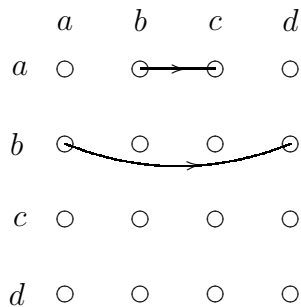
\begin{center}
\begin{tabular}{c}
\rxyg{0.7}{
,(10,-5);(15,-5)**\dir{-}?(.6)*\dir{>}
,(5,-10);(20,-10)**\crv{(12.5,-13)}?(.6)*\dir{>}
}  \\
\end{tabular}
\end{center}
This diagram corresponds to diagram 5 in \cite{4}.
It fails for it has no unbroken colour and is dynamically degenerate.
\vspace{1\baselineskip}

\newpage 

{\bf Diagram 2:} 

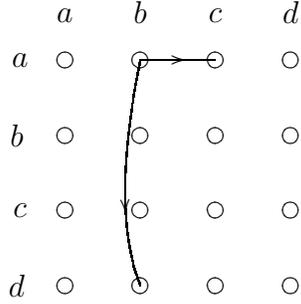
\begin{center}
\begin{tabular}{c}
\rxyg{0.7}{
,(10,-5);(15,-5)**\dir{-}?(.6)*\dir{>}
,(10,-5);(10,-20)**\crv{(8,-15)}?(.6)*\dir{>}
} \\ 
\end{tabular}
\end{center}
It corresponds to diagram 6 in \cite{4} 
and fails because it has no unbroken colours
and all the summands in the matrix algebra have to be 1-dimensional.
\vspace{1\baselineskip}

{\bf Diagram 3, 4:} 

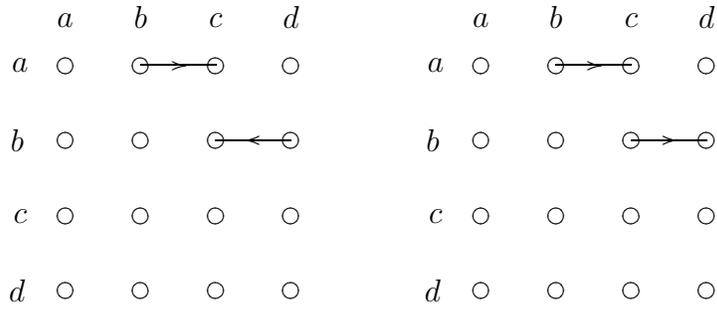
\begin{center}
\begin{tabular}{ccc}
\rxyg{0.7}{
,(10,-5);(15,-5)**\dir{-}?(.6)*\dir{>}
,(20,-10);(15,-10)**\dir{-}?(.6)*\dir{>}
}
&$\quad \quad$&
\rxyg{0.7}{
,(10,-5);(15,-5)**\dir{-}?(.6)*\dir{>}
,(15,-10);(20,-10)**\dir{-}?(.6)*\dir{>}
} \\
\end{tabular}
\end{center}
They correspond to  diagram 8 in \cite{4}. It has as
its corresponding model the electro-strong model which is treated
in detail  in \cite{4}.
\vspace{1\baselineskip}

{\bf Diagram 5:} 

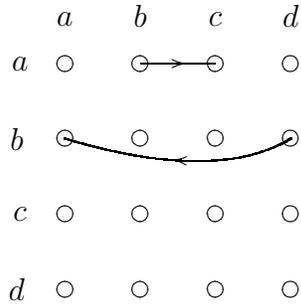
\begin{center}
\begin{tabular}{c}
\rxyg{0.7}{
,(10,-5);(15,-5)**\dir{-}?(.6)*\dir{>}
,(20,-10);(5,-10)**\crv{(15,-13)}?(.6)*\dir{>}
} \\
\end{tabular}
\end{center}
This diagram corresponds to diagram 20 in \cite{4}.
It  fails for it either exhibits a trivial little group or a charged neutrino.
\vspace{1\baselineskip}
\newpage
{\bf Diagram 6, 7:} 

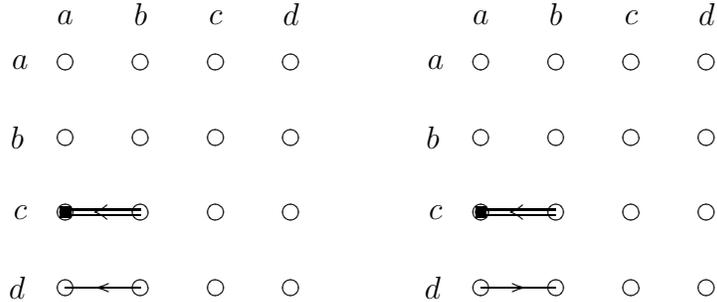
\begin{center}
\begin{tabular}{ccc}
\rxyg{0.7}{
,(5,-15)*\cir(0.4,0){}*\frm{*}
,(10,-15);(5,-15)**\dir2{-}?(.6)*\dir2{>}
,(10,-20);(5,-20)**\dir{-}?(.6)*\dir{>}
}
&$\quad \quad$&
\rxyg{0.7}{
,(5,-15)*\cir(0.4,0){}*\frm{*}
,(10,-15);(5,-15)**\dir2{-}?(.6)*\dir2{>}
,(5,-20);(10,-20)**\dir{-}?(.6)*\dir{>}
} \\
\end{tabular}
\end{center}
These models correspond to the diagrams
18 and 19 in \cite{4}. They reproduce the standard model of
particle physics with various possibilities for the colour groups
and certain sub-models.  
For the standard model algebra we find
\bb
\Aa_{SM} = \cc \oplus \hhh \oplus M_C (\cc) \oplus \cc,
\ee
where $C$ is the number of colours which has to be fixed by hand.
All the physical models produced by this diagram are treated in great detail in \cite{4}.
It is interesting to note that the maximal Dirac operator corresponding to the left
diagram (diagram 18 in \cite{4}) has the Standard Model Dirac operator as
its {\it maximal} Dirac operator. It does not contain any Lepto-Quarks!

\vspace{1\baselineskip}

\section{Conclusions}

This classification shows that the standard model takes an even more
prominent place among the finite spectral triples when passing from
$KO$-dimension zero to $KO$-dimension six, cutting down
the number of relevant Krajewski diagrams from 66 to seven.
It is most interesting to note that the Krajewski diagrams in $KO$-dimension
six form a proper sub-set of the Krajewski diagrams obtained in 
\cite{4} for $KO$-dimension zero. We could not show whether this
is a general feature or coincidence.  

Although one has to 
content oneself with a minimal version of the standard model, 
not allowing for massive neutrinos
in all generations and prohibiting  Majorana-masses for the right-handed neutrino 
and thus the 
see-saw-mechanism, this is  still consistent with experimental data. 
Furthermore the fermion-doubling problem is resolved,
as was shown in \cite{connes6,barrett6}. 

An important result is 
that a real, finite spectral triple in $KO$-dimension six allows a natural,
though ambiguous split of the Hilbert space into left-(right-)handed particle and antiparticle subspaces. 
One can therefore reduce the input for the spectral 
triple by the $S^0$-real structure which was needed 
in the case of $KO$-dimension zero.  This
reduction of input results in a  remarkable reduction of
the number of possible Krajewski diagrams and consequently a reduction
of the number of possible physical models. 

Extending the standard model by introducing massive right-handed
neutrinos, as done by Alain Connes \cite{connes6} and John Barrett
\cite{barrett6}, necessitates in a modification of the axioms of noncommutative
geometry, especially the orientability axiom \cite{connespriv}.

\vskip1cm
\noindent
{\bf Acknowledgements:} The authors would like to thank T. Sch\"ucker for careful proof
reading, Bruno Iochum and Thomas Krajewski for helpful discussions and
Alain Connes for insightful comments. 
We
gratefully acknowledge a scholarship of the Friedrich Ebert-Stiftung for J.-H. Jureit and a 
fellowship of the Alexander von Humboldt-Stiftung for C.A. Stephan.


\begin{thebibliography}{10}

\bibitem{1}
B.~Iochum, T.~Sch{\"u}cker, and C.~A. Stephan.
\newblock On a classification of irreducible almost commutative geometries.
\newblock {\em J. Math. Phys.}, 45:5003--5041, 2004.

\bibitem{2}
J.-H. Jureit and C.~A. Stephan.
\newblock On a classification of irreducible almost commutative geometries, a
  second helping.
\newblock {\em J. Math. Phys.}, 46:043512, 2005.

\bibitem{3}
T.~Sch{\"u}cker.
\newblock Krajewski diagrams and spin lifts.
\newblock 2005.

\bibitem{4}
J.-H. Jureit, T.~Sch{\"u}cker, and C.~A. Stephan.
\newblock On a classification of irreducible almost commutative geometries iii.
\newblock {\em J. Math. Phys.}, 46:072303, 2005.

\bibitem{Kraj}
T.~Krajewski.
\newblock Classification of finite spectral triples.
\newblock {\em J. Geom. Phys.}, 28:1--30, 1998.

\bibitem{connes6}
A.~Connes.
\newblock Noncommutative geometry and the standard model with neutrino mixing.
\newblock 2006.

\bibitem{barrett6}
J.~W. Barrett.
\newblock A lorentzian version of the non-commutative geometry of the standard
  model of particle physics.
\newblock 2006.

\bibitem{book}
A.~Connes.
\newblock {\em Noncommutative geometry}.
\newblock Academic Press, London and San Diego, 1994.

\bibitem{grav}
A.~Connes.
\newblock Gravity coupled with matter and the foundation of non-commutative
  geometry.
\newblock {\em Commun. Math. Phys.}, 182:155--176, 1996.

\bibitem{connespriv}
A.~Connes.
\newblock Private communication.

\bibitem{Pasch}
M.~Paschke and A.~Sitarz.
\newblock Discrete sprectral triples and their symmetries.
\newblock {\em J. Math. Phys.}, 39:6191--6205, 1998.

\bibitem{real}
A.~Connes.
\newblock Noncommutative geometry and reality.
\newblock {\em J. Math. Phys.}, 36:6194--6231, 1995.

\bibitem{algo}
J.-H. Jureit and C.~A. Stephan.
\newblock Finding the standard model of particle physics: A combinatorial
  problem.
\newblock {\em J. Comp. Phys.}, accepted for publication, 2005.

\bibitem{ko6}
C.A. Stephan.
\newblock  Almost-commutative geometry, massive neutrinos and the orientability axiom in KO-dimension 6.
\newblock hep-th/0610097, 2006
\end{thebibliography}
\end{document}